\documentclass[aps,pra,twocolumn,showpacs,preprintnumbers,amsmath,amssymb,
superscriptaddress]{revtex4}

\usepackage{mathptmx}  
\usepackage{dcolumn}                    
\usepackage{bm}                        
\usepackage{graphicx}
\usepackage{times}
\usepackage{epstopdf}
\usepackage{bbold}
\usepackage{xcolor}
\usepackage{natbib}
\usepackage{textcomp}
\usepackage[]{subfigure}

\begin{document}

\newcommand{\bfk}{\mathbf{k}}
\newcommand{\VK}{\mathbf{K}_0}
\newcommand{\Imag}{{\Im\mathrm{m}}}   
\newcommand{\Real}{{\mathrm{Re}}}   
\newcommand{\im}{\mathrm{i}}        
\newcommand{\talpha}{\tilde{\alpha}}
\newcommand{\ve}[1]{{\mathbf{#1}}}

\newcommand{\x}{\lambda}  
\newcommand{\y}{\rho}     
\newcommand{\T}{\mathrm{T}}   
\newcommand{\Pv}{\mathcal{P}} 
\newcommand{\vk}{\mathbf{k}} 
\newcommand{\vd}{\boldsymbol{d}} 
\newcommand{\vecp}{\boldsymbol{p}} 
\newcommand{\vecr}{\boldsymbol{r}} 

\newcommand{\ea}{\epsilon_k^A} 
\newcommand{\eb}{\epsilon_k^B} 

\newcommand{\N}{\underline{\mathcal{N}}} 
\newcommand{\Nt}{\underline{\tilde{\mathcal{N}}}} 
\newcommand{\g}{\underline{\gamma}} 
\newcommand{\gt}{\underline{\tilde{\gamma}}} 

\newcommand{\vq}{\ve{q}} 

\newcommand{\Tkp}[1]{T_{\vk\vp#1}}  
\newcommand{\muone}{\mu^{(1)}}      
\newcommand{\mutwo}{\mu^{(2)}}      
\newcommand{\epsk}{\varepsilon_\vk}
\newcommand{\epsp}{\varepsilon_\vp}
\newcommand{\e}[1]{\mathrm{e}^{#1}}
\newcommand{\dif}{\mathrm{d}} 
\newcommand{\diff}[2]{\frac{\dif #1}{\dif #2}}
\newcommand{\pdiff}[2]{\frac{\partial #1}{\partial #2}}
\newcommand{\mean}[1]{\langle#1\rangle}
\newcommand{\abs}[1]{|#1|}
\newcommand{\abss}[1]{|#1|^2}
\newcommand{\Sk}[1][\vk]{\ve{S}_{#1}}
\newcommand{\pauli}[1][\alpha\beta]{\boldsymbol{\sigma}_{#1}^{\vphantom{\dagger}}}

\newcommand{\eq}{Eq.}
\newcommand{\eqs}{Eqs.}
\newcommand{\cf}{\textit{cf. }}
\newcommand{\ie}{\textit{i.e. }}
\newcommand{\eg}{\textit{e.g. }}
\newcommand{\etal}{\emph{et al.}}
\def\i{\mathrm{i}}                            
\newcommand{\jacob}[1]{\textcolor{red}{#1}}

\title{Superfluid Breakdown and Multiple Roton Gaps in Spin-Orbit Coupled \\Bose-Einstein Condensates on an Optical Lattice}

\author{Daniele Toniolo}
\affiliation{Department of Physics, Norwegian University of
Science and Technology, N-7491 Trondheim, Norway}

\author{Jacob Linder}
\affiliation{Department of Physics, Norwegian University of
Science and Technology, N-7491 Trondheim, Norway}

\date{\today}


\begin{abstract}
We investigate the superfluid phases of a Rashba spin-orbit coupled 
Bose-Einstein condensate residing on a two dimensional square optical lattice in 
the presence of an effective Zeeman field $\Omega$. At a 
critical value $\Omega=\Omega_c$, the single-particle 
spectrum $ E_\bfk $ changes from having a set of four degenerate minima 
to a single minimum at $\vk=0$, corresponding to 	
condensation at finite or zero momentum, respectively. We describe this 
quantum phase transition and the symmetry breaking of the condensate phases. We use the 
Bogoliubov theory to treat the 
superfluid phases and determine the phase diagram, the excitation spectrum and the sound velocity 
of the phonon excitations. A novel dynamically unstable superfluid regime occurring when $\Omega$ 
is close to $\Omega_c$ is analytically identified and the behavior of the condensate quantum 
depletion is discussed. Moreover, we show that there are two types of roton excitations occurring in 
the $\Omega<\Omega_c$ regime and obtain explicit values for the corresponding energy gaps.
\end{abstract}
\maketitle
\textit{Introduction}. The recent realization of ultracold 
spin-orbit coupled (SOC) quantum gases \cite{lin_nature_11} has 
attracted high interest and resulted in considerable research efforts both 
on the theoretical and experimental side \cite{wang_prl_10, ho_prl_11, 
radic_pra_11, li_prl_12, wilson_prl_13}, in part due to the possibility to tune the 
spin-orbit interactions  \cite{zhang_scirep_13} in 
contrast to solid state materials. 
Ultracold quantum gases with spin-orbit coupling manifest novel types 
of superfluid and magnetic ground-states and have also been predicted to 
host topological excitations like
Majorana fermions \cite{zhang_prl_08}.

The SOC Bose-Einstein condensate (BEC) has intrinsic features that make it different 
from the standard BEC: the interaction among atoms make a SOC BEC stable 
since it cannot exist in the free regime \cite{ozawa:stability}, the SOC also breaks the Galileian invariance so that
the superfluid properties change in different reference frames 
\cite{zhu_epl_12}; for a review see \cite{galitski_spielman:review}.
Several works have considered different types of SOC in the continuous limit:
pure Rashba, mixed and symmetric Rashba-Dresselhaus, in two and three dimensions  
 \cite{zheng_jphysb_13, zhou_prl_13,martone_pra_12}. The exotic properties of the Mott insulating phase arising
from the superfluid-Mott insulator (SF-MI) transition \cite{cole_prl_12, sengupta_prb_12} were also considered in the case of an optically induced lattice.
However, an analytical quantitative description of the SF phase for a SOC BEC
in an optically induced lattice is still missing.

In this work, we consider a   
Bose-Einstein condensate with Rashba SOC residing on a 2D square optical lattice and prove that the SOC qualitatively affects 
the features of the superfluid phase. The system's parameters are the Zeeman-coupling $\Omega$, 
the strength of the spin-orbit coupling $\lambda$, the hopping $t$, and the intra- and interspecies interactions $U,U'$. We discuss the origin and magnitude of these terms in more detail later on. We will in this paper show three main results: 
I) with $ \lambda \gg t $ the existence of the SF is related 
to the ratio $ \Omega / U $ and not to $ t / U $ like in the usual 
Bose-Hubbard models; 
II) $\Omega$ can trigger a breakdown of 
SF in a window near the critical value $\Omega_c \equiv 2\lambda^2/t$, 
in this regime the excitation spectrum assumes complex values, indicating 
a dynamical instability toward a phase-separation \cite{law_prl_97};
 III) in the regime $ \Omega< \Omega_c $, the excitation spectrum has, besides 
the usual gapless phonon minimum localized at the condensation momentum, three gapped roton minima with different
gap energies $\Delta_\perp$ and $\Delta_\parallel$. We provide analytical evidences of all these results.

\textit{Bose-Hubbard formulation}.
It is possible to induce on a dilute atomic boson gas system, through laser-atom interactions,
a spin-momentum interaction such that the effective system has two coupled levels. In this sense 
one may speak of pseudospin-$\frac{1}{2}$ bosons. The confinement on a 2D plane and the 
periodic potential on it
can be experimentally realized through the action of counter-propagating lasers.
Our starting point is a two species Bose-Hubbard type Hamiltonian \cite{sengupta_prb_12}
   $ \mathcal{H} =\mathcal{H}_0 + \mathcal{H}_\text{int} $:
\begin{align}
\label{lattice ham}
 \mathcal{H}_0 &= \sum_{\langle i,j \rangle,\alpha\beta} 
[-t_\alpha b_{i\alpha}^\dag b_{j\alpha}\delta_{\alpha\beta} + 
\i\lambda b_{i\alpha}^\dag \hat{z}\cdot(\boldsymbol{\sigma} \times \vd_{ij} )_{\alpha\beta} b_{j\beta} ]\notag\\
& +\sum_{i\alpha\beta} [\delta b_{i\alpha}^\dag (\sigma_y)_{\alpha\beta} b_{i\beta} -
\Omega b_{i\alpha}^\dag (\sigma_z)_{\alpha\beta} b_{i\beta} - \mu b_{i\alpha}^\dag b_{i\alpha}\delta_{\alpha\beta}],\notag\\
 \mathcal{H}_\text{int} &=\sum_{i\alpha} \frac{U}{2} b_{i\alpha}^\dag b_{i\alpha}^\dag b_{i\alpha} b_{i\alpha} + 
  \sum_i U' b_{i A}^\dag b_{i B}^\dag b_{i A} b_{i B} 
\end{align}
Above, $i$ is the lattice site index, $\alpha$ and $\beta$ run over the two species
$A$,$B$, that correspond to the pseudospin $\pm \frac{1}{2}$,
$\mu$ is the chemical potential, $t_\alpha$ is the
hopping term, $ \lambda$ is the strength of the spin-orbit coupling,
$\hat{z}$ is the unit vector in $z$-direction,
$\vd_{ij}$ is the nearest neighbor (NN) vector between lattice sites
$i$ and $j$, $\boldsymbol{\sigma}$ is the Pauli matrix vector, $\delta$ 
is the detuning parameter, $ \Omega $ is the shift in chemical
potential due to the Zeeman interaction between spin and magnetic field. 
The square optical lattice is assumed to lie in the $xy$-plane. The interaction part 
$\mathcal{H}_\text{int}$ 
contains the intra- and interspecies interactions $U, U'$,  
we allow these coefficients to be different. We set $\hbar=1$ in what follows.
We diagonalize the non-interacting Hamiltonian $\mathcal{H}_0$ 
using the quasi-momentum basis $\{ b_{\bfk \alpha},b_{\bfk \alpha}^\dagger \}$: $b_{i\alpha} = \frac{1}{\sqrt{N_S}} \sum_\bfk b_{\bfk\alpha}\e{\i\vk\cdot\vecr_i}.$, $ N_s$ is the total number of sites. 
We focus on equal 
hopping coefficients $ t_A = t_B \equiv t $ 
and $ \delta = 0 $ for the sake of obtaining more tractable analytical expressions that allow for 
deeper physical insights. The energy bands are:
$E_{\bfk,\pm} =-2t(\cos k_x + \cos k_y) -\mu\pm \sqrt{\Omega^2 + 4\lambda^2 (\sin^2 k_x 
+\sin^2 k_y)}.$
The spectrum $ E_{\bfk,\pm} $ is invariant under parity 
$(k_x \rightarrow -k_x, k_y \rightarrow -k_y) $ and under 
permutation of $k_x$ and $k_y $, $(k_x \rightarrow k_y, k_y \rightarrow k_x) $, so the total 
symmetry group of $ E_{\bfk,\pm} $ is $ \mathbb{Z}_2 \otimes \mathbb{Z}_2 \otimes \mathbb{S}_2  $.
 The value of $ \Omega $ strongly affects the shape of $ E_{\bfk,-} $:
 with $ \Omega > \Omega_c \equiv 2 \lambda^2/t $ it has
 one minimum at $ (0,0) $; with $ \Omega < \Omega_c $ it has four degenerate
 minima at $ ( \pm k_0, \pm k_0 ) $,
\begin{align}
 {k_0}=\arcsin{ \sqrt{\big[1-\big(\Omega/\Omega_c\big)^2\big]\big/\big[1 + 
2\big(t/\lambda\big)^2\big]}}.
\end{align}
At the critical value $ \Omega = \Omega_c $,  it has one minimum at $ (0,0) $
 behaving as a fourth order power in momentum. We note that without a lattice structure the minima 
degeneracy 
 of $ E_{\bfk,-} $ is continuous in the $(k_x,k_y)$ plane, whereas it is discrete in our case so that
the SF phase is expected to be more robust towards quantum fluctuations. 
We define the operator basis $ \{d_{\bfk,-},d_{\bfk,+}\}$ that respectively 
annihilates a boson in the lower band $ E_{\bfk,-} $ and in the upper band  $ E_{\bfk,+} $. These 
are related to $\{ b_{\bfk A},b_{\bfk B}\}$ via the unitary matrix $\mathcal{P}$.
We are interested in a low-energy description of the system at $T=0$ and thus we consider
populated only the lowest energy band $E_{\bfk,-}$. This condition is
qualitatively satisfied taking $ \Omega > \max\{U,U'\} $. In fact, 
$ 2\Omega < \left(E_{\bfk,+} - E_{\bfk,-}\right) < 2\Omega + 8 \lambda $, and
$ \max\{U,U'\} $ is an estimate of the energy at disposal to scatter from 
the lower band to the upper band. We define $ E_{\bfk,-}\equiv E_\bfk $.
With this assumption $d_{\bfk,+} \to 0$ and the operators $b_{\bfk A}$ and $b_{\bfk B} $ 
are directly proportional to $d_{\bfk,-}\equiv d_{\bfk}$: 
$b_{\bfk A}=\alpha_\bfk d_\bfk$ and
$b_{\bfk B}=\beta_\bfk d_\bfk$, where we set $ \alpha_\bfk \equiv \mathcal{P}_{1,1}  $ and $\beta_\bfk \equiv
\mathcal{P}_{2,1} $. The
coefficients $\alpha_\bfk \in \mathbb{R} $  and ${\beta_\bfk} \in \mathbb{C}$  are the probability amplitudes for a particle
in the band $ E_\bfk $
to be of the $A$ or $B$ type. From the unitarity of $ \mathcal{P} $ 
it follows that $  \alpha (\bfk) ^2 + \left| \beta (\bfk) \right|^2 = 1 $;
\begin{align}
&\alpha_{\bf k}=
\sqrt{(1/2)\big[ 1+\big(1+\left(2 \lambda/\Omega\right)^2
(\sin^2 k_x + \sin^2 k_y)\big)^{-1/2}\big]} \notag\\
& \beta_{\bf k} =
\big[\big( \sin k_y -i \sin k_x\big)/\sqrt{\sin^2 k_x + \sin^2 k_y}\big] \sin\theta_{\bf k} 
\label{alphabeta}
\end{align}
We define $ \cos \theta_\bfk \equiv \alpha_\bfk $ for later purposes. The interaction Hamiltonian as a function of the operators $\{d_\bfk,d_\bfk^\dagger \}$ reads:
\begin{align}
\mathcal{H}_{int} &= \sum_{\bfk+\bfk'={\bf p}+{\bf p}'} \frac{U}{2 N_S} ( \alpha_\bfk \alpha_{\bfk'}
\alpha_{\bf p} \alpha_{\bf p'} + \beta_{\bfk}^* \beta_{\bf k'}^* \beta_{\bf p} \beta_{\bf p'})  d_\bfk^\dagger
d_{\bf k'}^\dagger d_{\bf p} d_{\bf p'}  \notag\\ 
&+ \sum_{\bf k+\bfk'={\bf p}+{\bf p'}} \frac{U'}{N_S} \alpha_\bfk \beta_{\bfk'}^* \alpha_{\bf p} \beta_{\bf p'}  d_\bfk^\dagger d_{\bfk'}^\dagger d_{\bf p} d_{\bf p'} 
\label{h interacting}
\end{align}
We note that the scattering coefficients in \eqref{h interacting} are invariant under parity.
We discard the upper energy band $ E_{\bfk,+} $ which corresponds to 
map the original $ 
\{A,B\} $ components into an effective one-component system with momentum-dependent interaction 
coefficients Eq. \eqref{h interacting}.

In the regime $ \Omega<\Omega_c$ the non-interacting energy spectrum $ E_\bfk $ has four 
degenerate minima which raises the issue of whether the condensation takes place at 
one or more momenta. As we discuss after the evaluation of the ground state energy \eqref{mean field 
hamiltonian}, 
the condensation momentum is unique when $U>U'$: this is the so called plane wave phase. Our analysis and results are restricted to this case. 

The shape of $E_\bfk$ changes varying $\Omega$ across 
$\Omega_c$, this determines a quantum phase transition. With $\Omega>\Omega_c$ the condensation momentum is
$ \VK = 0 $, the corresponding state preserves the parity symmetry in momentum space; 
with $\Omega<\Omega_c$ the condensation 
momentum is $\VK \neq 0$,  this is a symmetry broken phase because the corresponding condensate 
state breaks the parity symmetry in momentum space. A natural choice for the order parameter of 
this QPT is $ |\beta_{\VK}|^2 $ that passes from a non zero value with $\Omega<\Omega_c$ to zero 
with $\Omega>\Omega_c$, varying continuously.

\begin{figure}[th!]
     \centering
     \subfigure{%
         \includegraphics[width=0.48\linewidth]{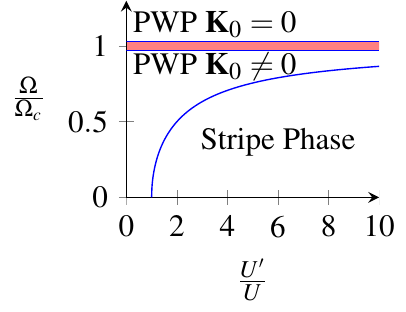}%
        \label{fig:phasediagram}%
        }%
     \subfigure{%
         \includegraphics[width=0.48\linewidth]{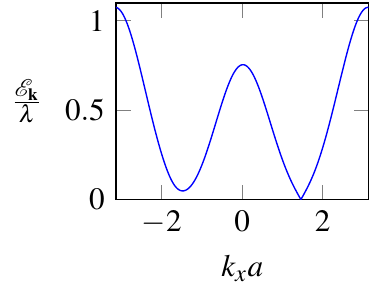}%
        \label{fig:phonon}%
        }%
    \caption{(Color online) \textit{Left panel:} The red (grey) stripe denotes the instability 
region. With 
$\Omega<\Omega_c$ and $ U'<U $ 
the stable phase is the plane wave phase (PWP) with finite condensation momentum. With $ U'>U $ the 
PWP and the striped phase are competing. With $\Omega>\Omega_c$ the favored phase is the PWP with 
condensation momentum equal to zero. Increasing $U'/U$ the phase boundary between the Striped 
Phase and the PWP tends to $1$. \textit{Right panel:} projection of the excitation spectrum 
$\mathcal{E}_\bfk$ on $ k_y=k_0$ showing the phonon excitation and roton gap. $\tilde{t}=0.08$, 
$\tilde{\Omega}=0.55$, $n=1$, $\tilde{U}=0.12$, $\tilde{U}'=0.11$.}
    \label{phasephonon}
\end{figure}

To treat the condensate phase we apply the Bogoliubov theory which is very well suited to capture the SF properties but not to investigate the SF-MI transition \cite{sengupta_prb_12}, the latter being outside the scope of the present work.  Let $\VK$ denote the condensation momentum which is zero or finite according to the value of $ \Omega $. We then have $d_{\VK}^\dag d_{\VK} = N_{\VK} \gg 1 $ and subsequently apply the Bogoliubov approximation $d_{\VK}^\dag \sim d_{\VK} \sim \sqrt{N_{\VK}}$.
We perform a mean-field approximation of Eq. \eqref{h interacting} by taking into account the  particle number fluctuations out of the condensate to the first order \cite{liu_pra_07}.
The final Hamiltonian is:
\begin{align}
&   \mathcal{H}=E_0 + \sum^{'}_\bfk \left( a_\bfk d_\bfk^\dagger d_\bfk + b_\bfk d_\bfk d_{2\VK -\bfk} +b_\bfk^*
                                        d_\bfk^\dagger d_{2\VK - \bfk}^\dagger \right)
\label{final hamiltonian}                                        
\end{align}
the symbol $ ' $ indicates that $ \VK $ is excluded from the sum. 
 With $ n= (N_A+N_B)/N_S $ we have:
\begin{align}
E_0&/N_S = n E_{\VK} + n^2\left[(U/2)\left(\alpha_{\VK}^4+|\beta_ {\VK }|^4 \right )+
                                           U'\alpha_{\VK}^2 |\beta_{\VK}|^2\right]  \notag\\
a_\bfk& = E_\bfk - E_{\VK} + n U [2\alpha_{\bfk}^2 \alpha_{\VK}^2 +
                 2|\beta_\bfk|^2|\beta_{\VK}|^2 -\alpha_{\VK}^4 -|\beta_{\VK}|^4 ] \notag\\ 
+&n U' [\alpha_\bfk^2 |\beta_{\VK}|^2+\alpha_{\VK}^2(|\beta_{\bfk}|^2 -2 |\beta_{\VK}|^2)+ 2 \alpha_\bfk \alpha_{\VK}
              \Re(\beta_\bfk \beta_{\VK}^*) ]\notag\\
b_\bfk&= (n/2)  U (\alpha_{\VK}^2 \alpha_\bfk \alpha_{2\VK-\bfk} +
           \beta_{\VK}^{*2}\beta_\bfk \beta_{2\VK-\bfk}) \notag\\ &+ nU'\alpha_{\VK}\beta_{\VK}^*
            					( \alpha_\bfk \beta_{2\VK-\bfk}+\alpha_{2\VK-\bfk} \beta_\bfk )   
\label{mean field hamiltonian}
\end{align}
$E_0 $ is the ground state energy. Considering $ \Omega < \Omega_c $ we can compare $ E_0 $ 
with the ground state energy obtained by supposing that the condensate state is equally populated by 
atoms with momenta $ \VK $ and $-\VK $ (striped phase), this is obtained taking into account in 
the interaction Hamiltonian Eq. (\ref{h interacting}) values of the momenta $\{\bfk, \bfk', \bf{p}, 
\bf{p}'\}$ equal to $\{\pm \VK\ ,\pm \VK\ ,\pm \VK\ ,\pm \VK\ \}$, $\{\pm \VK ,\mp 
\VK , \pm \VK, \mp \VK \}$, or $\{\pm \VK ,\mp 
\VK , \mp \VK, \pm \VK \}$. With $ U>U'$ the favored phase is the plane wave phase whereas 
with $U'>U$ the boundary between the two phases is
\begin{align}
\Omega/\Omega_c=\sqrt{2}t/\sqrt{((x+1)/(x-1))(\lambda^2+2t^2)-\lambda^2},
\end{align}
with $ x=U'/U$ (see Fig. \ref{phasephonon}). We have checked that possible condensate phases 
that populate \eg all the four minima of $ E_\bfk $ always have a higher ground state energy.

We diagonalize the mean field Hamiltonian Eq. \eqref{final hamiltonian} making sure to preserve the boson commutation relations \cite{tsallis}, obtaining the excitation spectrum and the final Hamiltonian:
\begin{align}
\label{excitation_spectrum}
&\mathcal{E}_\bfk=\frac{1}{2} \left(a_\bfk-a_{2\VK-\bf k}+ \sqrt{(a_\bfk +a_{2\VK-\bfk})^2 - 16 
|b_\bfk|^2} \right) \\
\label{ham c operators}
&\mathcal{H}=E_0 + \frac{1}{2} \sum^{'}_\bfk ( \mathcal{E}_\bfk - a_\bfk) + \sum^{'}_\bfk
 \mathcal{E}_\bfk C_\bfk^\dagger C_\bfk
\end{align}
$C_\bfk , C_\bfk^\dagger$ are the bosonic annihilation and creation
operators of the excitations $ \mathcal{E}_\bfk $. In the non-interacting limit $ U=U'=0 $, we have $ b_\bfk=0 $ and 
$ a_\bfk=E_{\bfk}-E_{\VK} $,
so $ \mathcal{E}_\bfk $ reduces to $E_{\bfk}-E_{\VK}  $. We see that 
$ \mathcal{E}_{\bf K_0}=0 $ so that the excitation spectrum is gapless at the condensation 
momentum, moreover the square root term of Eq. \eqref{excitation_spectrum}, which is 
responsible for the phonon excitations, has reflection symmetry across $ \VK $. Eq. 
\eqref{excitation_spectrum} is the general form of the excitation energies. Before to analyze the features of the excitation spectrum \eqref{excitation_spectrum} in the two 
regimes $\Omega<\Omega_c$ and $\Omega>\Omega_c$, we determine the effective mass of the particles of 
our model and also the values of $\lambda$, $\Omega$, $t$, $U$, $U'$ that place our system in the
SF phase.

\textit{Effective masses, superfluidity criterium}.
The effective masses are the eigenvalues of $ \partial^2 E|_{\VK} $; with $ \Omega < \Omega_c $, this matrix is non-diagonal so that the effective masses correspond to motion along a rotated set of orthogonal axis $x'$ and $y'$. These are:
\begin{align}
 m^*_\pm= 2\big[ (t^3/\lambda^2) \sin k_0 \tan k_0 [(1+(\lambda/t)^2)\pm 1 ]\big]^{-1}
\label{eff masses}
\end{align}
To give a physical interpretation of Eq. \eqref{eff masses} we normalize each quantity choosing $ 
\lambda$ as the unit of energy and consider the cases $ \tilde t \equiv t/\lambda \gg 1 $, $ 
\tilde t \ll 1 $. Summarizing:
\begin{align}
\text{For } \tilde t \gg 1:&\text{ } m^*_-=\tilde t / R, m^*_+=1/(\tilde t R),\notag\\
\text{For } \tilde t \ll 1:&\text{ } m^*_-=\tilde \Omega / R, m^*_+=\tilde \Omega/R 
\end{align}
with $ \tilde \Omega= \Omega/\lambda $ and $ R = \big(1-(\Omega/\Omega_c)^2\big) $. The 
criterium that we use in order to determine the parameter values ensuring that our system is in a SF 
phase, and not in a MI one, comes from the one-component Bose-Hubbard model. There, $ m^* \sim 1/t $ 
and the superfluidity is ensured with $ m^* U < 1$ \cite{greiner}. Considering the same condition $ 
m^*_\pm U < 1$ we see that with  $ \tilde t \ll 1 $ the parameter guiding the SF is $ \Omega $ and 
not $ t $. Moreover, we see that with $ \Omega \to \Omega_c^{-} $ the SF is always strongly disfavored.
With $ \Omega > \Omega_c$, $E_\bfk $ has only one minimum in (0,0) and the effective mass is isotropic 
$m^*= [2t(1-\Omega_c/\Omega)]^{-1}$. With $ \lambda \rightarrow 0$ $(\Omega_c \rightarrow 0)$, 
this reduces to the usual result 
for the standard Bose-Hubbard model $ m^* = \frac{1}{2t} $; also in this case with $ \Omega \to 
\Omega_c^{+} $ SF is disfavored. 

The general formula for the sound velocity from Eq. \eqref{excitation_spectrum} is:
\begin{align}
 c_{x,\pm}  
 = \partial_{k_x} a_\bfk \big|_{\VK} \pm 
  \sqrt{a_{\VK}\left[ \partial_{k_x}^2 \left(a_\bfk -2 |b_{\bfk}|\right)
                                              \big|_{\VK}\right] }
\label{sound velocity}
\end{align}
It can be shown that if $ \partial_{k_x}^2 \left(a_\bfk -2 |b_{\bfk}|\right) \big|_{\VK}<0$, $ 
\mathcal{E}_\bfk $ becomes complex around $ \VK $, so looking at the 
sound velocity is a natural tool to find possible instabilities of $ \mathcal{E}_\bfk $.

\begin{figure}[h!]
     \centering
      \subfigure{%
         \includegraphics[height=0.36\linewidth]{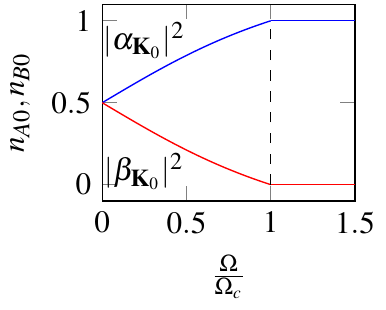}%
         \label{components}%
         }%
      \subfigure{%
         \includegraphics[height=0.36\linewidth]{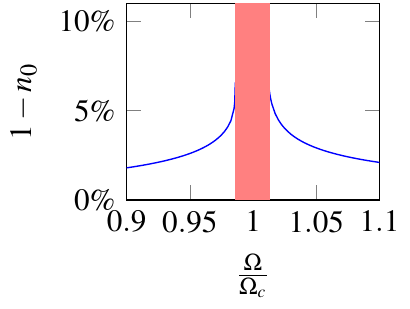}%
         \label{depletion}%
         }%
    \caption{(Color online) \textit{Left panel:} the relative population of A and B atoms in the 
condensate phase: 
$n_{A0}\equiv|\alpha_{{\bf K}_0}|^2$, $n_{B0}\equiv|\beta_{{\bf K}_0}|^2$. The dashed vertical line 
corresponds to the value of $\Omega_c$, with $\Omega>\Omega_c$, $n_{A0}=1$, $n_{B0}=0$.  
\textit{Right panel:} quantum 
depletion (percentage of the total particle number); the depletion grows approaching the 
instability region, red (grey), but it is 
nevertheless small. The parameters of 
the plot are: $\tilde{t}=1$,  
$\tilde{\Omega}_c=2$, $n=1$, $\tilde{U}=0.1$, $\tilde{U}'=0.05$, $N_s=10^4$. }
    \label{compdep}
\end{figure}

\textit{$\Omega>\Omega_c$, excitation spectrum, sound velocity, instability.}
In this case $ E_\bfk $ features only a minimum at $\vk=0$, so that $\VK=0$ and $\beta_{\bf 0} =0$. Then:
\begin{align}
\mathcal{E}_\bfk = \sqrt{ \big[E_\bfk-E_{\VK} + nU (2\alpha_\bfk^2-1) 
+nU'(1-\alpha_\bfk^2)\big]^2- n^2 U^2 \alpha_\bfk^4 } \notag
\label{one minimum}
\end{align}
A phonon excitation appears in the limit $\vk\to0$ with sound velocity $c\equiv c_x=c_y$: 
\begin{align}
c=\sqrt{2nU\big[t - 2(\lambda^2/\Omega)-n(\lambda/\Omega)^2\left(U-U'\right)\big]}.
\end{align}
When $ \lambda \rightarrow 0 $, $ c \to \sqrt{2nUt} $ that is the one-component Bose-Hubbard result for $c$.
Approaching the critical value $ \Omega \rightarrow \Omega_c^{+} $, both $ \mathcal{E}_\bfk $ and $ c $ 
become imaginary under the condition:
$ n(U-U')/2\Omega>\left(\Omega/\Omega_c-1\right) $, 
this is one of our main results. The imaginary eigenvalues are indicative of a novel dynamical instability for the superfluid phase on an optical lattice when including SOC.
A physical interpretation of this instability is related to the real 
underlying two component $\{A,B\}$ system that seems to enter a phase-separation 
regime \cite{law_prl_97}. This can be understood by considering the left panel of Fig. \ref{compdep} where we plot the relative population of the atomic species A (spin up) and B (spin down) in 
the condensate. Due to the Zeeman coupling, Eq. \eqref{lattice ham}, the atoms of the species A are 
energetically favored respect to the species B in the condensed phase. The two atomic species 
coexist in the condensate until $\Omega$ reaches the value $\Omega_c$ at which point the species B 
is expelled 
from the BEC.  

\textit{ $ \Omega < \Omega_c $, sound velocities, instability, roton excitation.} In this case $ 
E_{\bfk} $ has four degenerate minima localized at $ ( \pm k_0, \pm k_0 ) $;
without loss of generality we assume that the condensation momentum is equal to $\VK = (k_0,k_0)$. The excitation spectrum has a cusp at $\VK$, proving the existence of phonons. The slope differs slightly on the positive and negative direction of the $ k_x $, respectively $ k_y $, axis, this is associated with the anisotropy of the effective masses. The sound velocity $ c_{x,\pm} = c_{y,\pm} $  is given in the footnote \cite{cvelocity}, its structure 
is in agreement with 
\cite{martone_pra_12} that considered the continuum case (no optical lattice). 
From the explicit analytical form of the sound velocity it is possible to determine the values of $ 
\Omega $  such that $ c_{x,\pm} $   becomes complex and the 
excitation spectrum becomes dynamically unstable. We 
consider two regimes $ \tilde t > 1 $ and $ \tilde t < 1 $:
\begin{align}
&\tilde t > 1: \tilde \Omega_c(1-\tilde t^{-2}/8)<\tilde \Omega<\tilde \Omega_c ,\notag\\
&\tilde t < 1: \tilde \Omega_c\big(1- n(\tilde t/8) (\tilde U - \tilde U^{'})\big)<\tilde 
\Omega<\tilde \Omega_c
\end{align}

Thus, just as in the case $ \Omega>\Omega_c $, an instability appears when $\Omega$ is close to the 
critical value $\Omega_c$.
In addition to the phonon minimum occuring at the condensation momentum, a peculiar feature resulting from the presence of spin-orbit coupling is the presence of additional roton minima. Such objects are absent in multicomponent Bose-Einstein condensates without spin-orbit interactions and may be understood as a consequence of the degenerate nature of the minima in the excitation spectrum $E_\bfk$ without interactions. 
We find that the roton gaps are \textit{not} degenerate in 
spite of the single-particle spectrum minima being degenerate. The excitation spectrum Eq. 
(\ref{excitation_spectrum}) has the usual phonon minimum localized at $ {\VK} $ whereas we find that 
the positions of the roton minima are close to the positions of the degenerate minima of 
the single-particle spectrum as long as one considers weak interaction parameters $U$,$U'$. In fact 
discarding the second order terms in $ U $ and $ U' $, Eq.\eqref{excitation_spectrum} approximately reduces to $a_\bfk$ far from the 
condensation momentum $\VK $. With $\VK = 
(k_0,k_0)$, the positions of the roton excitations are then: $ (k_0,-k_0), (-k_0,k_0), (-k_0,-k_0) 
$. The roton gaps $\Delta(\vk)$ are:
\begin{align}
&\Delta_\perp \equiv \Delta(k_0,-k_0) = \Delta(-k_0,k_0) = n U \left( 2 \alpha_{\VK}^4 -2 \alpha_{\VK}^2 +1 \right)\notag\\
&\Delta_\parallel \equiv \Delta(-k_0,-k_0) = nU -n(U+U')2 \alpha_{\VK}^2 \left(1-\alpha_{\VK}^2 \right).
\end{align}
All gaps are always positive as long as $ U> U' $, which is the regime we are considering 
(plane-wave phase). As seen, there exist two types of gaps $\Delta_\perp$ and $\Delta_\parallel$: 
one gap for the roton excitations closest to the condensation momentum ($\Delta_\perp$) and one gap 
for the roton excitation farthest away from it $(\Delta_\parallel)$. The degeneracy of the minima in the non-interacting case is 
partially lifted when adding interactions $U$ and $U'$.

\textit{Quantum depletion}. The BEC depletion at a temperature $T$ is the
average relative
number of particles not belonging to the BEC:
$ 1-n_0=(1/N) \sum_{{\bf k} \neq {\VK}} \langle d_{\bf 
k} ^\dagger d_{\bf k} \rangle $,
the operators $d_\bfk $ as in Eq. \eqref{mean field hamiltonian}, $ n_0 \equiv  \langle 
d_{\VK} ^\dagger d_{\VK} \rangle / N $.
At $T=0$ only the quantum fluctuations contribute to the depletion. Performing a basis change from 
$d_\bfk$ 
to the quasiparticle operators $C_\bfk$ (see \eg section 4 in Ref. \cite{tsallis}) it allows to 
obtain:  $ 1-n_0 = \sum_{{\bf k} \neq {\VK}} 
 1/2N\big(|a_{\bf k}+a_{2\VK-\bfk}|/\sqrt{(a_\bfk 
+a_{2\VK-\bfk})^2 - 16 
|b_\bfk|^2} -1 \big) $. Inside the instability region, the above expression of the 
quantum 
depletion loses its meaning because the sum above becomes complex. In the right panel of Fig \ref{compdep}, we 
present a numerical evaluation of the quantum depletion, the depletion increases slightly upon approaching the 
dynamical unstable region but nevertheless remains small for a system of finite size, we also 
numerically evaluate the depletion 
as a function of $t/\lambda$  with $\Omega$ closed to the instability region, both on the left and 
right side, and found that it is always lesser than $10\%$. In the thermodynamic limit, the BEC does 
not exist at the edges of the instability region but the quantum depletion rapidly decreases in the 
neighborhood of the edges in such a way the instability  region is still 
well defined.

\textit{Summary}. In summary, we have established a phase-diagram for the superfluid state of a 
SOC BEC in the presence of a 2D square optical lattice. We have 
identified an instability regime in a window of values for the Zeeman-coupling $\Omega$ near a 
critical value $\Omega_c $ where the excitation energies become complex. We have also derived 
analytical expressions for the roton excitations appearing in the system, and shown that there are 
two types of inequivalent roton gaps.

\textit{Acknowledgments}. D.T. and J.L. were supported by the Research Council of Norway through Grant 216700.

\end{document}